\documentclass[pra,onecolumn,superscriptaddress]{revtex4-1}

\usepackage{amssymb}
\usepackage{graphicx}
\usepackage{bbm,amsmath,amssymb}
\usepackage{epsfig}
\usepackage{array}
\usepackage{dsfont}

\newcommand{\ket}[1]{\ensuremath{\left|{#1}\right\rangle}}

\begin{document}

\title{Quantum Simulator for Transport Phenomena in Fluid Flows}

\author{A. Mezzacapo}
\email{ant.mezzacapo@gmail.com}
\address{Department of Physical Chemistry, University of the Basque Country
UPV/EHU, Apartado 644, E-48080 Bilbao, Spain}

\author{M. Sanz}
\address{Department of Physical Chemistry, University of the Basque Country
UPV/EHU, Apartado 644, E-48080 Bilbao, Spain}

\author{L. Lamata}
\address{Department of Physical Chemistry, University of the Basque Country
UPV/EHU, Apartado 644, E-48080 Bilbao, Spain}

\author{I. L. Egusquiza}
\address{Department of Theoretical Physics and History of Science,
University of the Basque Country UPV/EHU, Apartado 644, E-48080 Bilbao, Spain}

\author{S. Succi}
\address{Istituto per le Applicazioni del Calcolo ``M. Picone'' CNR, I-00185 Rome, Italy}
\address{Institute for Applied Computational Science, Harvard University, Oxford Street, 33, 02138 Cambridge, USA} 

\author{E. Solano}
\address{Department of Physical Chemistry, University of the Basque Country
UPV/EHU, Apartado 644, E-48080 Bilbao, Spain}
\address{IKERBASQUE, Basque Foundation for Science, Maria Diaz de Haro 3, 48013 Bilbao, Spain}

\begin{abstract}
Transport phenomena still stand as one of the most challenging 
problems in computational physics. 
By exploiting the analogies between Dirac and lattice Boltzmann 
equations, we develop a quantum simulator based on pseudospin-boson 
quantum systems, which is suitable for encoding fluid dynamics 
transport phenomena within a lattice kinetic formalism. 
It is shown that both the streaming and collision processes of 
lattice Boltzmann dynamics can be implemented with controlled quantum operations, using a heralded quantum protocol to encode non-unitary scattering processes. 
The proposed simulator is amenable to realization in controlled quantum platforms, such as ion-trap quantum computers or circuit quantum electrodynamics processors. 
\end{abstract}

\date{\today}
\maketitle

Transport phenomena in fluid flows play a crucial role for many applications in science and engineering.
Indeed, a large variety of natural and industrial processes depend critically on the transport of mass, momentum and energy of chemical species by means of fluid flows across material media of assorted nature \cite{TRA}. The numerical simulation  of such transport phenomena still presents a major challenge to modern computational fluid dynamics. Among the reasons for this complexity stand out the presence of strong heterogeneities and huge scale separation in the basic mechanisms, namely advection, diffusion and chemical reactions \cite{ADR,KAX}. In the last two decades, a novel concept for the solution of transport phenomena in fluid flows has emerged in the form of a minimal lattice Boltzmann (LB) kinetic equation. This approach is based on the statistical viewpoint typical of kinetic theory \cite{Benzi92,AIDUN}. LB is currently used across a broad range of problems in fluid dynamics, from fully developed turbulence in complex geometries to micro and nanofluidics \cite{GENOA,EPL}, all the way down to lattice gas automata~\cite{MacNamara} and quark-gluon applications~\cite{PRLMiller}.

Recent improvements in ion trap and superconducting circuit experiments make these platforms ideal for challenging quantum information and simulation tasks. Trapped-ion experiments have demonstrated  quantum information and simulation capabilities~\cite{Lanyon14,Nigg14,Lanyon11}, including the quantum simulation of highly correlated fermionic systems~\cite{Casanova12}, fermionic-bosonic models~\cite{Mezzacapo12,Stojanovic12} and lattice gauge theories~\cite{Hauke13}. Superconducting circuit setups can host nowadays top-end quantum information protocols, such as quantum teleportation~\cite{Steffen13} and topological phase transitions~\cite{Roushan14}. These quantum devices are approaching the complexity required to simulate both classical and quantum nontrivial problems, as proposed by Feynman some decades ago ~\cite{Feynman82}.  Efforts in designing quantum algorithms for the implementation of fluid dynamics make use of quantum computer networks~\cite{Yepez01,Pravia03}. In these works, the quantum degrees of freedom are used on the same ground as classical parameters, and the exponential gain of quantum computers is not properly exploited. In contrast, systems described by pseudospins coupled to bosonic modes, such as the aforementioned ion-trap and superconducting circuit platforms, can enjoy quantum superposition and have advantages with respect to pure-qubit quantum computers in simulating fluids.

In this article, we propose a quantum simulation of lattice Boltzmann dynamics, using coupled pseudospin-boson quantum platforms. Based on previously established analogies between Dirac and LB equations, we define here a full quantum mapping of transport equations in fluid flows. The LB dynamics is simulated sequentially by performing particle streaming and collision steps. The non-unitary collision process can be implemented with an heralded protocol, by sequential collapses of an ancillary qubit. 
The proposed mapping is amenable to realization in trapped-ion and superconducting circuit platforms.

\section*{Results}

The lattice Boltzmann equation is a minimally discretized version of the original Boltzmann's kinetic equation, in which the fluid is modeled as an ensemble of particles that move and collide within a uniform lattice. The lattice Boltzmann dynamics is described by the equation
\begin{equation}
\label{QLB}
(\partial_t  + v^b_i \nabla_b )f_i(\vec{x},t) = -A_{ij}[f_j(\vec{x},t)-f^{eq}_j(\vec{x},t)].
\end{equation}
Here, $f_i(\vec{x},t)$ is the $i$th component of the particle fluid density associated with the lattice site $\vec{x}$ at the time $t$, and with discrete velocity $\vec{v}_i$. The macroscopic fluid density at the site $\vec{x}$ is retrieved as $\rho(\vec{x},t)=\sum_if_i(\vec{x},t)$, while the fluid velocity is defined as the weighted sum of the discrete velocities, $\vec{u}(\vec{x},t)=1/\rho \sum_i f_i(\vec{x},t)\vec{v}_i$. 
The velocity components $f_i\vec{v}_i$, with $i=1,2,...Q$, satisfy mass-momentum-energy conservation laws and rotational symmetry. Typical lattices are {\it D2Q9} or {\it D3Q15} models, for the case of two dimensions with $9$ speeds, and three dimensions with $15$ speeds, respectively \cite{QIAN}.

Collisional properties are here expressed in scattering-relaxation form, making use of the local equilibrium distribution $f_i^{eq}(\vec{x},t)$. The LB approach to compute the dynamics associated with Eq.~(\ref{QLB}) uses sequential computational steps. One initially performs a displacement (free-streaming) of each distribution component $f_i(\vec{x})$ towards the nearest-neighbor lattice site 
pointed at by the discrete velocity  $\vec{v}_i$. From there, the equilibrium distribution function $f_i^{eq}(\vec{x},t)$ is computed and the outcome of the collisional process is retrieved. Further iterations of these calculations allow the propagation of the lattice dynamics in time. We address the question of whether all these steps can be performed in a quantum simulator with practical quantum computing protocols.

The formal analogy between the Dirac and LB equations was first highlighted in~\cite{Benzi92,Succi93}, where the velocity distribution of the particle 
is treated in a similar fashion as a relativistic spinor. This analogy is further exploited in the Majorana representation of the Dirac equation, by using real spinors~\cite{Fillion13}. The Dirac (Majorana) equation reads ($\hbar=1$ here and in the following)
\begin{equation}
\label{Majorana}
i\partial_t \Psi_i + i\alpha^b_{ij}\nabla_b \Psi_j=\beta_{ij}\Psi_j,
\end{equation}
where we have defined the Dirac (Majorana) streaming matrices $\alpha_{ij}^b$, 
mass term $\beta_{ij}$, and the imaginary prefactor $i$ proper of quantum mechanical evolution. 

Notice that the streaming matrices of the LB equation are diagonal, while the $\alpha_{ij}$, which generate a Clifford algebra, cannot be simultaneously diagonalized. Additionally, the mass matrix $\beta_{ij}$ is Hermitian, while standard collision matrices come in real symmetric form in the LB equation. Therefore, a complete codification of the LB scheme in quantum language requires the implementation of diagonal streaming matrices and of purely imaginary symmetric scattering matrices.

\begin{figure}[t] 
\includegraphics[scale=0.16]{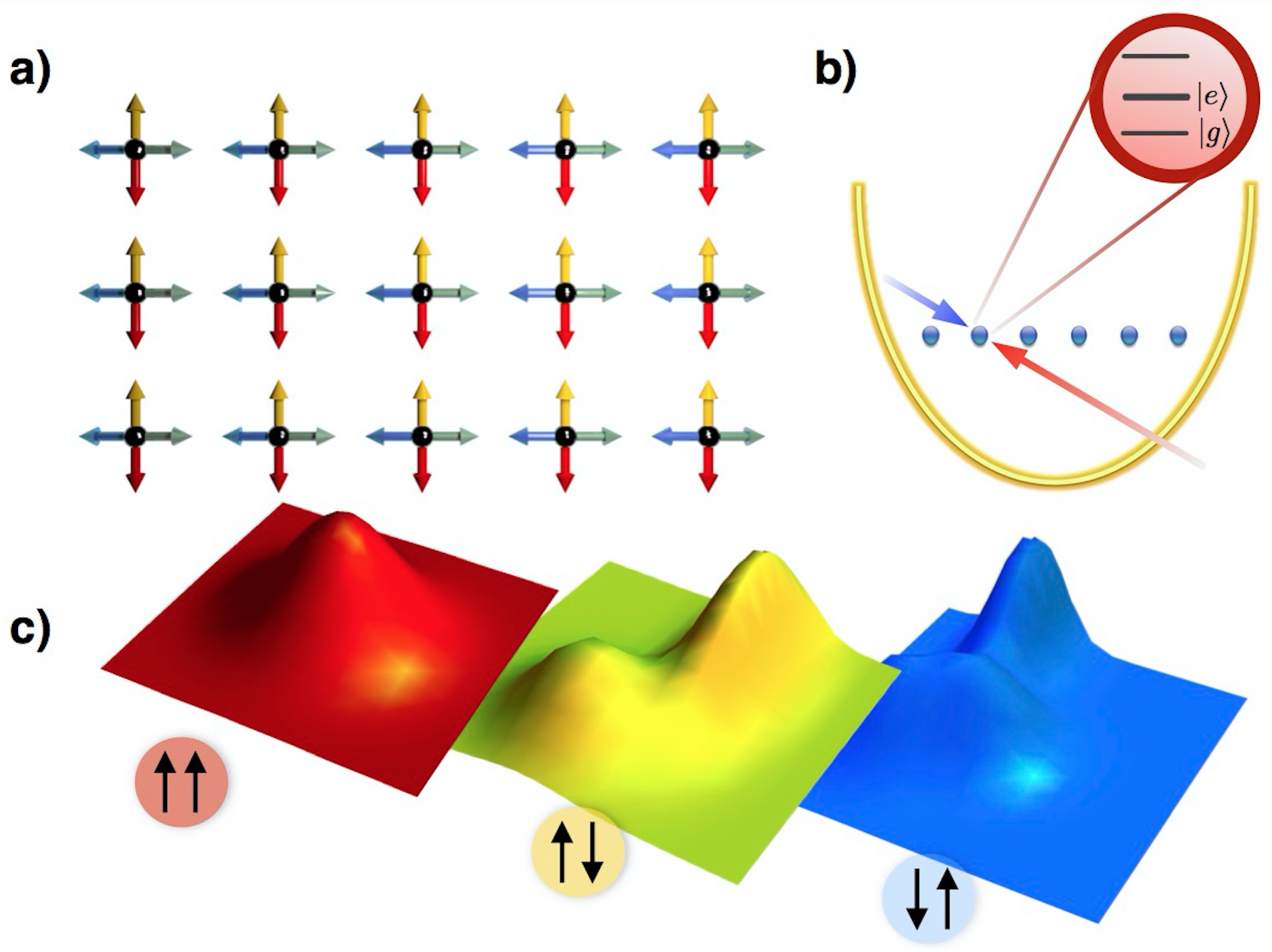}
\caption{(color online) (a) The distribution of the fluid density on a 2-dimensional lattice can be simulated, for example, via normal motional modes and internal levels of a set of trapped ions (b). (c) Superposition of two motional modes entangled with pseudo spin states can encode velocity distributions in different lattice directions. }
             \label{LBPhaseSpace}
\end{figure}

The components of the fluid density distribution function $f_i(\vec{x},t)$ can be encoded in a set of quantum states $\ket{\Psi_i}$ defined on a proper Fock space. For example, in two dimensions, the distribution of the fluid density over the two coordinates can be described by a real quantum wavefunction that encodes the state of two bosonic modes, as depicted in Fig.~\ref{LBPhaseSpace}. In the $x$-quadrature representation, it reads $\ket{\Psi_i}=\int dx_1 dx_2 f_i(x_1,x_2)\ket{x_1}\ket{x_2}$, where $f_i(x_1,x_2)$ is a real distribution and $\ket{x_{1(2)}}$ the eigenstate of the quadrature of the first (second) bosonic mode. Several quantum distributions $\ket{\Psi_i}$ can be used by entangling the bosonic state to a multi-level system, such as a set of pseudospins, therefore the state of the complete system is given by $\ket{\Psi}=\sum_i\eta_i\ket{i}\otimes\ket{\Psi_i}$, with $\eta_i$ being real-valued coefficients. In order to keep a real-valued representation of $\ket{\Psi}$, to be identified with a fluid density distribution function, one has to act only with purely imaginary interaction matrices.

The quantum simulation of the Dirac equation was originally proposed \cite{Lamata07} and afterwards realized in a trapped-ion experiment~\cite{Gerritsma10}. In general, streaming interactions involving matrices in the Dirac or Majorana representation $\alpha_{ij}^b \nabla_b$ can be implemented by using a pair of pseudospins coupled to one or more bosonic modes. In terms of creation and annihilation operators $a_b(a_b^{\dag})$ for the bosonic mode in the $b$ direction, one can then consider $p_b=i\nabla_b=i(a_b-a_b^{\dag}) $ and write Eq.~(\ref{Majorana}) on the pseudospin-bosonic Hilbert space of $\ket{\Psi}$,
\begin{equation}
i\partial_t \ket{\Psi(t)}  =  K_b\alpha^b i(a_b-a_b^{\dag}) \ket{\Psi(t)} +\beta\ket{\Psi(t)},
\end{equation} 
where $K_b$ stands for the pseudospin-boson coupling and $\alpha^b$ act upon the pseudospin degrees of freedom.

Thus, the three streaming matrices $\alpha_{ij}^b$ are written in the Dirac representation as $\alpha^b=-\sigma^x_1\otimes \sigma_2^b$, in a pseudospin representation and the diagonal mass term as $\beta={\sigma}^z_1\mathds{I}_2$. These streaming matrices are diagonalized via the operators $S_b=1/\sqrt{2}(\beta+\alpha^b)$~\cite{Fillion13}, which have to be physically implemented as quantum gates.  
Defining $S_b=\exp(-iH_bt)$, the associated generators read $H_b=A{\sigma}^z_1\otimes\mathds{I}_2+B\sigma_1^x\otimes\sigma_2^b$, with $A=\frac{\sqrt{2}\pi}{4}$ and $B=\frac{\pi}{2\sqrt{2}}$. In this way, a purely imaginary streaming step $i\beta\nabla_b$ can be built, which mimics the diagonal streaming of the LB equation. The total wavefunction after the streaming steps can be retrieved with a sequential implementation, following the operator splitting method~\cite{Succi93}. For example, in a 2-dimensional lattice, one has  
\begin{equation}
\label{StreamingCollision}
\ket{\Psi(t_{n+1})}=(S_y^{-1}D_yS_y)(S_x^{-1}D_xS_x)C\ket{\Psi(t_n)}.
\end{equation}
The last collision step $C$, which scrambles particle distributions in different directions, is discussed below.

Standard collision operators in LB theory are represented by real symmetric 
matrices associated with non-unitary evolution operators. 
On the other hand, typical controlled quantum mechanics experiments produce unitary dynamics. 
Nevertheless, one can probabilistically encode non-unitary dynamics in a quantum device with a heralded protocol, by performing controlled operations conditioned on the state of an ancillary qubit, and then using the state of the latter as a flag for the success of the protocol.
We consider a purely imaginary symmetric scattering matrix $\Omega$, whose quantum evolution equation  
reads $i\partial_t \Psi_i=\Omega_{ij}\Psi_j$, providing a non-unitary evolution 
operator that describes lattice collisions $C=\exp(-i\Omega \Delta t)$. 

The collision operator can be decomposed in a weighted sum of two commuting unitary operators, $C=U_\alpha+\gamma U_\beta$, with the constraint $||C||\leq 1+\gamma$, assuming without loss of generality that $\gamma >0$.    

\begin{figure}[t] 
\includegraphics[scale=0.15]{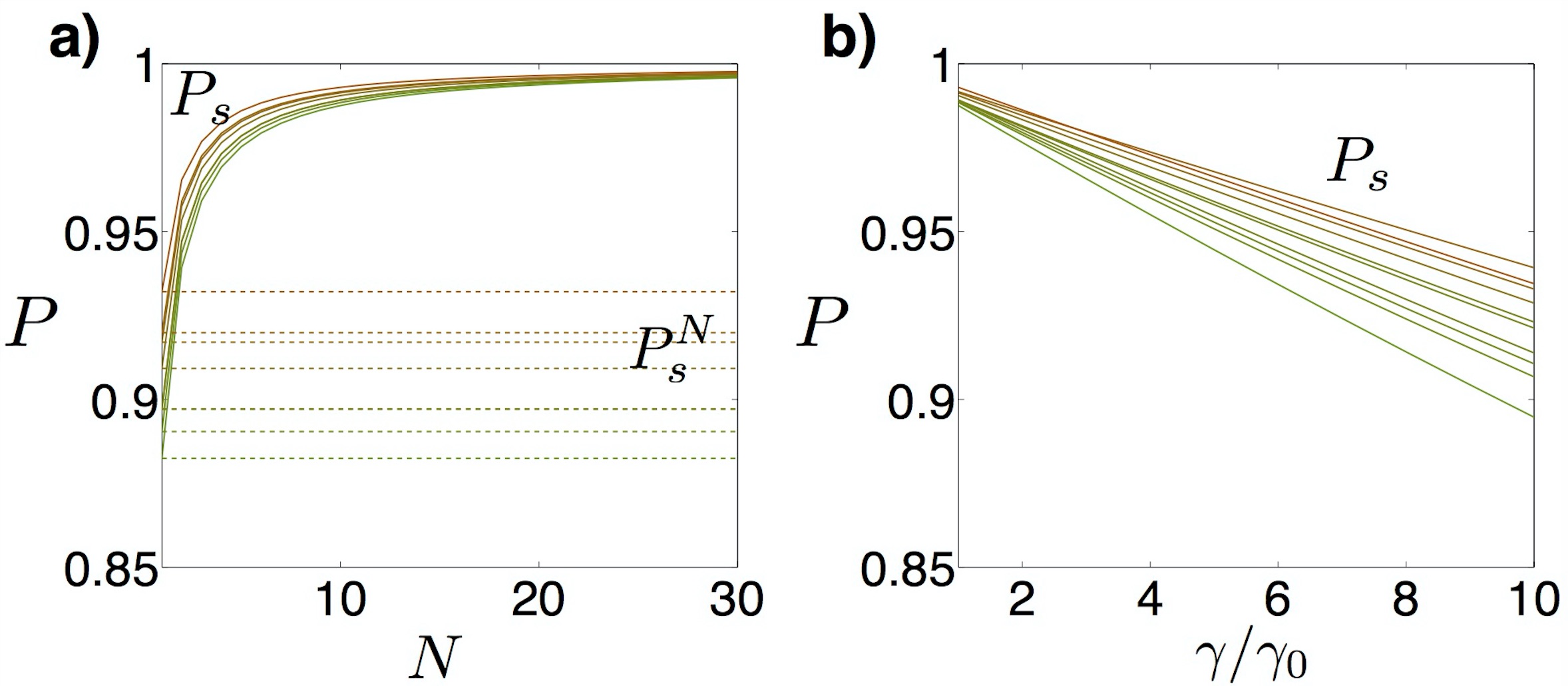}
\caption{(Color online) Probabilities of success $P$ for the implementation of the scattering process. (a)~Probability of success $P_s$ per time step of simulating real symmetric random matrices as a function of the number of ancillary measurements $N$ (solid lines), together with accumulated probabilities for the whole protocol $P_s^N$(dashed lines). Each curve represents a different instance of a random matrix. (b)~ Probability of success of a single step as a function of $\gamma/\gamma_0$, when $N=10$.}
             \label{SuccessP}
\end{figure}

Given a specific diagonalizable collision operator $C$ and weight $\gamma$, one can then find 
its decomposition in terms of unitaries. 
In order to find a decomposition in terms of unitaries, $C$ must first be diagonalized as $C=VDV^\dagger$. 
This reduces the problem of finding $U_\alpha$ and $U_\beta$ down to an eigenvalue 
equation, $\delta_i=\alpha_{i}+\gamma\beta_{i}$, with
$\delta_i$, $\alpha_i$ and $\beta_i$ being the $i$th eigenvalues of 
the collision and unitary operators respectively. 
Notice that, due to the properties of the scattering matrix, $\delta_i\in\mathds{R}^+$. 
Taking into account the normalization conditions, one has the system of equations
\begin{equation}
\begin{cases}
\delta_i=\alpha_{i}+\gamma\beta_{i}\\
|\alpha_{i}|=1\\
|\beta_{i}|=1.
\end{cases}
\end{equation}
The eigenvalues $\alpha_i$, $\beta_i$ can now be written as a function of the initial collision operator and weight $\gamma$,
\begin{eqnarray}
\label{alphaparts}
\textrm{Re}(\alpha_{i})&=& \frac{\delta_i ^2-\gamma ^2+1}{2 \delta_i }\nonumber\\
\textrm{Im}(\alpha_{i})&=& \frac{\sqrt{-\delta_i ^4+2 \delta_i ^2 \left(\gamma ^2+1\right)-\left(\gamma
   ^2-1\right)^2}}{2 \delta_i }\nonumber\\
   \textrm{Re}(\beta_{i})&=& \frac{\delta_i ^2+\gamma ^2-1}{2 \delta_i  \gamma }\nonumber\\
   \textrm{Im}(\beta_{i})&=& -\frac{\sqrt{-\delta_i ^4+2 \delta_i ^2 \left(\gamma
   ^2+1\right)-\left(\gamma^2-1\right)^2}}{2 \delta_i  \gamma }.
\end{eqnarray}
The unitary operators $U_{\alpha (\beta)}$ are reconstructed via $\left (U_{\alpha(\beta)}\right )_{ij}=V_{in}^\dagger \alpha_n (\beta_n)V_{nj}$.
The real domain of Eqs.~(\ref{alphaparts}) provides the range of validity of the 
method developed here. 
Simple algebra leads to the set of inequalities  
\begin{equation}
\label{GammaBounds}
|-1+\delta_i|\leq\gamma\leq1+\delta_i ,\; \; \forall i.
\end{equation}
By defining $\delta_M$ and $\delta_m$ as the maximal and minimal eigenvalues of the spectrum of $C$, the system of inequalities in Eq.~(\ref{GammaBounds}) can be reduced to one of the two inequalities $|-1+\delta_m|\leq\gamma\leq1+\delta_m$ or $|-1+\delta_M|\leq\gamma\leq1+\delta_m$, respectively when $|-1+\delta_m|\leq|-1+\delta_M|$ or $|-1+\delta_m|\geq|-1+\delta_M|$. If longer evolution times $t$ are considered, the spectral range of $C$ changes accordingly. The weighted $\gamma$-sum derived here can be implemented with quantum computing algorithms, using ancillary qubits and controlled $U_\alpha$ and $U_\beta$ gates~\cite{Wiebe12}. By measuring the ancilla state, one can determine whether the desired operation has been performed or not. The success of the protocol depends on the weighted sum of unitary operators, with a failure probability $P_f= \gamma ||U_\alpha - U_\beta||^2/(\gamma+1)^2$.   

As $P_f$ is an increasing function of $\gamma$, choosing $\gamma_0=\min\{|-1+\delta_m|,|-1+\delta_M|\}$ maximizes the probability of success. This directly connects the simulation time of the scattering process $C$ with the best choice for $\gamma$. To propagate the dynamics of a given collision process $C$, one can split the step time $\Delta t$ into $N$ time intervals $\Delta t/N$ and perform the heralded protocol at each step, such that $C=\exp(-i\Omega_{ij}\Delta t/N)^N$. At each step, one has a collision operator $\exp(-i\Omega_{ij}\Delta t/N)$, with an optimal $\gamma_{0}$. In this way, as the step size gets smaller, the success probabilities for each step increase, while the total success probability accumulates  single success rates from the individual steps. In Fig.~\ref{SuccessP}a, we plot the success probability $P_s(N)=1-P_f(N)$ of the simulation of the single step, as a function of $N$, for random symmetric purely imaginary matrices. As expected, the success probability per step increases as the size for the single time step gets smaller. The success of the whole protocol $P_s^N$ is constant and does not depend on $N$. In Fig.~\ref{SuccessP}b is shown that the optimal protocol is performed at $\gamma=\gamma_0$.

\section*{Discussion}
The scheme proposed can be adapted to a variety of transport fluid problems. As an example, we consider the implementation of an advection-diffusion process in two spatial dimensions. The dynamics of the transported species, e.g. pollutants or bacteria, is described by the equation
\begin{equation}
\label{advection}
\partial_t \rho + \nabla \cdot (\rho \vec{U}) = D \Delta \rho,
\end{equation}
where $\rho =\sum_{i=1}^4 f_i$ is 
the scalar field transported by a fluid with space-dependent 
velocity $\vec{U} = (U_x,U_y)$ and constant diffusivity $D$. 

The problem in Eq.~(\ref{advection}) can be recast in LB form, as in Eq.~(\ref{QLB}). The corresponding equilibrium distribution function is defined as
\begin{equation}
\label{feq}
f_i^{eq} = w_i \left[\rho + \frac{\rho \vec{U} \cdot \vec{c}_i}{c_s^2} \right],
\end{equation}
with $w_i=1/4$, $c_s^2 = 1/2$.
Note that, by definition, the space-time dependence of the local equilibria
is entirely carried by the macroscopic fields $\rho$ and $\vec{U}$.

The scattering matrix reads
$A_{ij} = \sum_{k=1}^4 A_i^{(k)} \omega_k A_j^{(k)}$, where $A_i^{(1)} = 1_i \equiv (1,1,1,1)$,
$A_i^{(2)} = c_{ix} \equiv (1,0,-1,0)$, $A_i^{(3)} = c_{iy} \equiv (0,1,0,-1)$ and $A_i^{(4)} = c_{ix}^2-c_s^2 \equiv (1/2,0,1/2,0)$ are the four eigenvectors. 

The first three corresponding eigenvalues are given by 
\begin{equation}
\label{omegavalues}
\omega_1=0, \; \; \omega_2=\omega_3=\frac{1}{1/2 + D/c_s^2} \, ,
\end{equation}
which follows from mass conservation and the expression of the diffusion constant $D = c_s^2 (1/\omega-1/2)$, respectively. 
By choosing different values for $\omega_2$ and $\omega_3$, one
can implement anisotropic diffusivities along the $x$ and $y$ directions.
Finally, $\omega_4$ dictates the decay of higher order fields 
of no direct macroscopic significance, hence it is chosen as 
$\omega_4=1$ so as to annihilate the corresponding field in a single
collision step. 

The relative scattering matrix $\Omega_{ij}$ is defined by $-A_{ij} [f_j(\vec{x};t)-f_j^{eq}(\vec{x};t)]\equiv\Omega_{ij} f_j(\vec{x},t)$. 
By choosing a Couette flow, e.g. 
$U=U_0(y,0)$, where $L$ is the typical size of the fluid domain, one 
has $f^{eq}_i=w_i\rho(1+u_i)$, with $u_1=-u_3=U_0 y/c_s^2$ and $u_2=u_4=0$. 
Here, velocities are numbered $1 \div 4$ counterclockwise starting from the $+x$ direction. 

\begin{figure}[t] 
\includegraphics[scale=0.23]{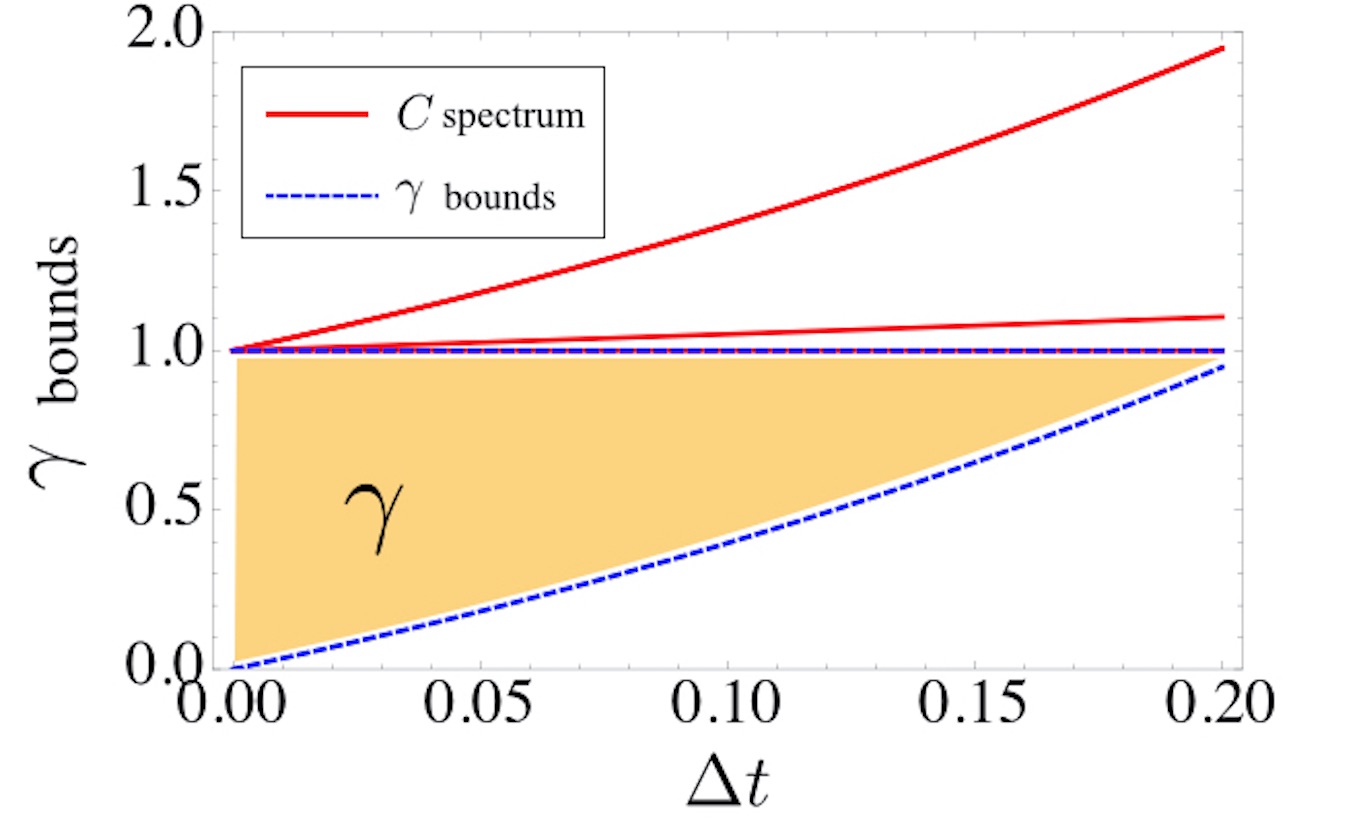}
\caption{(Color online)  Spectrum of a collision operator (solid red line) for advection-diffusion process of a four-speed lattice as a function of the evolution time step $\Delta t$, in units of $1/\omega_4$. The allowed region for $\gamma$ is bounded by dashed blue lines using Eq.~(\ref{GammaBounds}) and shadowed in the picture.}   
             \label{Gamma4Speed}
\end{figure}

The latter defines the quantum scattering matrix as composed of three contributions, namely  $i \Omega_{ij}f_j=-i A_{ij}[f_j+ w_j \rho + w_j \rho u_j]$, where $u_1=-u_3 \equiv U_0/c_s^2 (a_y+a^\dag_y)$ is proportional to the position quadrature of the bosonic mode associated with the $y$ direction. The three contributions to the scattering matrix represent classical linear wave propagation and damping, mass conservation and macroscopic advection, respectively. They can be implemented with the quantum simulation protocol previously introduced. The bounds to $\gamma$ can be obtained, e.g., for the first contribution 
to the scattering matrix $-A_{ij}$, by computing the spectrum of $C=e^{-A\Delta t}$ for different time steps $\Delta t$, for $D=0.05$, in units of $1/\omega_4$. The result is shown in Fig.~\ref{Gamma4Speed}.

Natural quantum platforms for prospective implementation of the proposed scheme could be ions trapped in linear Paul traps or superconducting circuit setups, in which the sequential streaming and collision steps in Eq.~(\ref{StreamingCollision}) can be realized. The pseudospin-bosonic state can be encoded, in the case of ion traps, in the internal level and motion modes of the ions~\cite{Haeffner08}, while in a superconducting architectures, one can use the first levels of charge-like qubits, e.g. transmon qubits, and microwave resonators~\cite{Devoret}.  One may consider opening similar avenues in other quantum technologies as is the case of quantum photonics~\cite{Szameit} and Bose-Einstein condensates~\cite{Weitz}. 

A practical implementation of the protocol proposed can make use of many-body interactions, involving couplings with bosonic modes. These type of gates have been considered in superconducting architectures~\cite{MezzacapoManyBody} or in ion-trap platforms~\cite{MullerManyBody}. For a four-speed lattice, the diagonal streaming processes can be realized with a combination of a qubit-boson interaction and two entangling gates among the qubits. For example, the corresponding evolution operator for the streaming in the $X$ direction can be written as 
\begin{equation}
\label{MSEq}
U_x=\exp\left[\phi\alpha^1(a_1-a_1^\dag)\right]=R_1^{z\dag}(\pi/4)R_2^{y\dag}(-\pi/4)U^\dag_C(\pi/4)\exp\left[\phi\sigma_1^x(a_x-a_x^\dag)\right]U_C(\pi/4)R_1^{z}(\pi/4)R_2^{y}(-\pi/4),
\end{equation}
where we have defined an entangling operation between the two qubits $U_C=\exp\left[-i(\pi/4)\sigma_1^z\sigma_2^z\right]$ and local rotations of the $i$-th qubit about the $j$-th axis, $R_i^j(\theta)=\exp(-i\theta\sigma_i^j)$. The $U_x$ interaction can then be diagonalized in the qubit space via the realization of two $S_x$ matrix, $S^\dag_xU_xS_x$, which can be achieved by a combination of entangling two-qubit gates and a phase gate. In the case of more internal degrees of freedom of the lattice, the two-body entangling gate can be substituted by a collective interaction $U_C\rightarrow\exp\left[-i(\pi/4)\sum_{i<j}\sigma_i^z\sigma_j^z\right]$. Similar reasoning applies to the streaming in the $Y$ direction, considering a different bosonic mode $a_y$. The unitary matrices that implement the collision process $U_{a(b)}=\exp(-iH_{a(b)}t)$ can be implemented in a controlled way~\cite{Wiebe12} by using an additional ancillary qubit $\Psi_A$ and performing the quantum gate $\exp\left[-i(\sigma_A^z\pm\mathds{1}_A)\otimes H_{a(b)}t\right]$. These gates can be decomposed in general with a Lie-Trotter-Suzuki decomposition, in terms of many-body interactions of the type $\sigma_A^{i_\alpha}\sigma_1^{j_\alpha}\cdots\sigma_N^{k_\alpha}$, $(\sigma_A^z\pm\mathds{1}_A)\otimes H_{a(b)}=\sum_\alpha  \eta_\alpha\sigma_A^{i_\alpha}\sigma_1^{j_\alpha}\cdots\sigma_N^{k_\alpha}$, with $\{i_\alpha,j_\alpha...k_\alpha\}\in\{x,y,z\}$. These many-body interactions can be obtained by sequential implementation of collective gates and single qubit rotations~\cite{MezzacapoManyBody,MullerManyBody}. In the case of a Couette flow, the term with the linear spatial dependence of the scattering matrix can be implemented by considering a single qubit rotation entangled with a bosonic displacement, similar to Eq.~(\ref{MSEq}). The quantum resources necessary to implement Lie-Trotter-Suzuki decompositions scale polynomially in the internal degrees of freedom of the lattice, and sub-polynomially in the digital error~\cite{Berry07}.  Notice also that the quantum resources needed are invariant with respect to the size of the simulated lattice. The latter will depend on the accessibility and readability of distributions over Fock spaces in practical implementations, e.g. the ability to characterize distributions over current quadrature in superconducting architectures~\cite{LehnertPRL,Lehnertarxiv}. 

Note that the above scheme readily extends to the case of reactive flow, by augmenting the collision operator with a local source term proportional to the chemical reaction rate. Such kind of advection-diffusion-reaction phenomena in complex geometries, say catalytic reactors, represent a very active area of applications of the LB scheme. Further developments may include the implementation of hydrodynamic non-linearities to model the Navier-Stokes fluid dynamic equations. This requires the inclusion of quadratic terms in the LB equilibrium distribution. Such nonlinear behavior can be provided in a quantum mechanical experiment by preparing multiple copies of the system~\cite{Leyton}, feedback 
mechanisms~\cite{Ringbauer}, or non-unitary operations induced by measurements.

We have developed a protocol to reproduce the dynamics of fluid transport phenomena in a quantum mechanical experiment, by using pseudospins coupled to bosonic modes that can be implemented in different quantum platforms. This proposal paves the way to quantum simulation and retrieval of complex classical fluid dynamics in controlled quantum systems. 

\section*{Author Contributions}
A.M. performed the calculations and the numerical analysis. 
M. S. and A. M. did the scheme in Fig.1. A. M., M. S., L. L., I. L. E. 
and E. S. contributed to the design of the quantum algorithm. 
S. S. contributed to the design and implementation of the 
algorithm for transport phenomena. 
All authors contributed to the writing of the manuscript.  

\section*{Acknowledgments}
We acknowledge financial support from Basque Government Grants IT472-10 and IT559-10, Spanish MINECO FIS2012-36673-C03-02, Ram\'on y Cajal Grant RYC-2012-11391, UPV/EHU Project No. EHUA14/04, UPV/EHU UFI 11/55, PROMISCE and SCALEQIT European projects.

\section*{Additional Information}
The authors declare no competing financial interests.

\end{document}